\documentclass{ws-procs975x65}

\begin{document}

\title{Traversable wormholes supported by dark gravity}

\author{Francisco S.~N.~Lobo}
\address{Centro de Astronomia e Astrof\'{\i}sica da Universidade de Lisboa,\\ Campo Grande, Ed. C8 1749-016 Lisboa, Portugal\\
E-mail: flobo@cii.fc.ul.pt}

\begin{abstract}

A fundamental property in wormhole physics is the flaring-out condition of the throat, which through the Einstein field equation entails the violation of the null energy condition. In the context of modified theories of gravity, it has also been shown that the normal matter can be imposed to satisfy the energy conditions, and it is the higher order curvature terms, interpreted as a gravitational fluid, that sustain these non-standard wormhole geometries, fundamentally different from their counterparts in general relativity. We review recent work in wormhole physics in the context of modified theories of gravity.

\end{abstract}


\bodymatter

\section{Introduction}

Wormholes are shortcuts in spacetime and are extremely useful as a theoretician's probe of the foundations of general relativity. A fundamental property is the flaring-out condition of the throat, which through the Einstein field equation entails the violation of the null energy condition (NEC) \cite{Morris:1988cz}. Note that the NEC violation involves a stress-energy tensor such that $T_{\mu\nu}k^\mu k^\nu <0$, where $k^\mu$ is any null vector \cite{Morris:1988cz}. Matter that satisfies this property has been denoted exotic matter. Thus, as the violation of the energy conditions is a problematic issue, it is useful to minimize its usage. For instance, one may obtain this in the context of thin-shell wormholes, by using the cut-and-paste procedure \cite{WHthinshell,WHthinshell2}. In this context, the exotic matter is concentrated at the throat of the wormhole, which is localised on the thin shell. 

Another approach lies within modified gravity, where normal matter is imposed to satisfy the energy conditions, and it is the higher order curvature terms that support these exotic spacetimes. In the context of modified gravity, the gravitational field equation may be written as $G_{\mu\nu}\equiv R_{\mu\nu}-\frac{1}{2}R\,g_{\mu\nu}= \kappa^2 T^{{\rm eff}}_{\mu\nu}$, where $\kappa^2 =8\pi G$, and for notational simplicity we consider $\kappa^2=1$; $ T^{{\rm eff}}_{\mu\nu}$ is the effective stress-energy tensor. Thus, in modified gravity, it is the effective stress-energy tensor involving higher order derivatives that is responsible for the violation of the NEC, i.e., $T^{{\rm eff}}_{\mu\nu} k^\mu k^\nu < 0$. This approach has been explored in $f(R)$ gravity \cite{modgravity1}, curvature-matter couplings \cite{modgravity2a,modgravity2b}, conformal Weyl gravity \cite{modgravity3} and in braneworlds \cite{modgravity4}, amongst other contexts. In this 
paper, we review the $f(R)$ gravity and conformal Weyl gravity scenarios.

\section{Wormholes in modified theories of gravity}

\subsection{$f(R)$ modified theories of gravity}

In this section, we consider $f(R)$ modified theories of gravity, which has recently been extensively explored as a possible cause of the late-time cosmic acceleration (see \cite{Lobo:2008sg} for a review). The action is given by
\begin{equation}
S=\frac{1}{2\kappa^2}\int d^4x\sqrt{-g}\;f(R)+\int d^4x\sqrt{-g}\;{\cal L}_m(g_{\mu\nu},\psi)
\,,
\end{equation}
where ${\cal L}_m$ is the matter Lagrangian density, in which matter is minimally coupled to the metric $g_{\mu\nu}$ and $\psi$ collectively denotes the matter fields.

Varying the action with respect to the metric, one deduces the gravitational field equation written in the following form $G_{\mu\nu}= T^{{\rm eff}}_{\mu\nu}$, as mentioned in the Introduction, where the effective stress-energy tensor, $T^{{\rm eff}}_{\mu\nu}$, is given by
\begin{eqnarray}
T^{{\rm eff}}_{\mu\nu}= \frac{1}{F}\left[T^{(m)}_{\mu\nu}+\nabla_\mu \nabla_\nu F
-\frac{1}{4}g_{\mu\nu}\left(RF+\nabla^\alpha \nabla_\alpha F+T\right) \right]    \,,
    \label{gravfluid}
\end{eqnarray}
where $F=df/dR$.

The flaring out condition for wormhole geometries imposes $T^{{\rm eff}}_{\mu\nu} \, k^\mu k^\nu< 0$, which yields the following generic condition in $f(R)$ gravity:
\begin{eqnarray}
\frac{1}{F}\left(T^{(m)}_{\mu\nu}\, k^\mu k^\nu +\, k^\mu k^\nu \nabla_\mu \nabla_\nu F 
\right) < 0   \,.
    \label{NECgravfluid}
\end{eqnarray}
This condition has been extensively explored\cite{modgravity1}, and specific wormhole solutions have been found. In principle, one may impose the condition $T_{\mu\nu} k^\mu k^\nu\ge 0$ for the normal matter threading the wormhole. Note that in general relativity, i.e., $f(R)=R$, we regain the condition for the matter stress-energy tensor violation of the NEC, i.e., $T_{\mu\nu} \, k^\mu k^\nu< 0$.

Thus, in the context of $f(R)$ modified theories of gravity it is the higher order curvature terms, interpreted as a gravitational fluid, that sustain these non-standard wormhole geometries, fundamentally different from their counterparts in general relativity. 

\subsection{Wormhole geometries in conformal Weyl gravity}

An intriguing modified theory of gravity is conformal Weyl gravity, involving the following purely gravitational sector of the action
\begin{eqnarray}\label{action1}
I_W&=&-\alpha \int d^4x
\sqrt{-g}\;C_{\mu\nu\alpha\beta}\,C^{\mu\nu\alpha\beta}
 \,,
\end{eqnarray}
where $C_{\mu\nu\alpha\beta}$ is the Weyl tensor, and $\alpha$ is
a dimensionless gravitational coupling constant. It was argued that in analogy to the principle of local gauge invariance that severely restricts the structure of possible Lorentz invariant actions in flat spacetimes, then the principle of local conformal invariance is a requisite invariance principle in curved spacetimes \cite{Kazanas:1988qa}. 

To this effect, one may rewrite the gravitational field equation as $G_{\mu\nu}=\frac{1}{4\alpha}T^{{\rm eff}}_{\mu\nu}$. Note that this relationship differs fundamentally from the Einstein field equation, as one is considering a dimensionless gravitational coupling constant $\alpha$, contrary to the Newtonian gravitational constant $G$. Nevertheless, the gravitational field equation written in this form proves extremely useful in deducing a definition of the NEC, in terms of the effective stress energy tensor, from the Raychaudhuri expansion
term $R_{\mu\nu}k^\mu k^\nu$.

Now, the effective stress energy tensor is given by $T^{{\rm eff}}_{\mu\nu}=T^{(m)}_{\mu\nu}+T^{(W)}_{\mu\nu}$. The first term, $T^{(m)}_{\mu\nu}$, in the effective stress energy tensor, is defined in terms of the matter stress energy tensor, $T_{\mu\nu}$, and is given by $T^{(m)}_{\mu\nu}\equiv \frac{3}{2R}\,T_{\mu\nu}$, where $R$ is the curvature scalar. The second term $T^{(W)}_{\mu\nu}$ may be denoted as the curvature Weyl stress energy tensor, and is provided by $T^{(W)}_{\mu\nu}\equiv -\frac{6\alpha}{R}\,\overline{W}_{\mu\nu}$, with the tensor $\overline{W}_{\mu\nu}$ defined as
\begin{eqnarray}
\overline{W}_{\mu\nu}&=&-\frac{1}{6}g_{\mu\nu}R^{;\beta}{}_{;\beta}
+R_{\mu\nu}{}^{;\beta}{}_{;\beta}
-R_{\mu}{}^{\beta}{}_{;\nu\beta}-R_{\nu}{}^{\beta}{}_{;\mu\beta}
    \nonumber    \\
&&-2R_{\mu\beta}R_{\nu}{}^{\beta}+\frac{1}{2}g_{\mu\nu}
R_{\alpha\beta}R^{\alpha\beta}+\frac{2}{3}R_{;\mu\nu}
+\frac{1}{6}g_{\mu\nu} R^2\,.
    \label{Weylmod}
\end{eqnarray}
Through the the Bianchi identities and the conservation of the stress energy tensor $T^{\mu\nu}{}_{;\nu}=0$, one verifies the following conservation law $T^{(W)\mu\nu}{}_{;\nu}=\frac{3}{2R^2}\, T^{\mu\nu}R_{,\nu}$.

The violation of the NEC, $T^{\rm eff}_{\mu\nu} k^\mu k^\nu <0$, imposes the following generic condition in conformal Weyl gravity
\begin{eqnarray}
\frac{1}{6R}\,T_{\mu\nu}k^\mu k^\nu -\frac{\alpha}{R}\left[R_{\mu\nu}{}^{;\beta}{}_{;\beta} -R_{\mu}{}^{\beta}{}_{;\nu\beta}-R_{\nu}{}^{\beta}{}_{;\mu\beta} -2R_{\mu\beta}R_{\nu}{}^{\beta}+\frac{2}{3}R_{;\mu\nu}  \right] k^\mu k^\nu < 0
\,.
    \label{NECWeylmod}
\end{eqnarray}
In this context, we refer the reader to a plethora of solutions found in \cite{modgravity3}.

\section{Conclusion}

In this work, we have explored the possibility that wormholes be supported by modified theories of gravity. We imposed that the matter threading the wormhole satisfies the energy conditions, and it is the higher order curvature derivative terms that support these nonstandard wormhole geometries, fundamentally different from their counterparts in general relativity.

\section*{Acknowledgements}
FSNL acknowledges financial support of the Funda\c{c}\~{a}o para a Ci\^{e}ncia e Tecnologia through the grants CERN/FP/123615/2011 and CERN/FP/123618/2011.


\begin{thebibliography}{99}


\bibitem{Morris:1988cz}
  M.~S.~Morris and K.~S.~Thorne,
  Am. J. Phys.  {\bf 56}, 395 (1988).

\bibitem{WHthinshell} 
  M.~Visser,
  Phys.\ Rev.\ D {\bf 39}, 3182 (1989);
%
\bibitem{WHthinshell2}
 N.~M.~Garcia, F.~S.~N.~Lobo and M.~Visser,
  Phys.\ Rev.\ D {\bf 86}, 044026 (2012).  

\bibitem{modgravity1}
  F.~S.~N.~Lobo, M.~A.~Oliveira,
  Phys.\ Rev.\ D  {\bf 80}, 104012 (2009).

\bibitem{modgravity2a}
 N.~M.~Garcia, F.~S.~N.~Lobo,
  Phys.\ Rev.\ D {\bf 82}, 104018 (2010).

\bibitem{modgravity2b}
N.~M.~Garcia, F.~S.~N.~Lobo,
  Class.\ Quant.\ Grav.\ {\bf 28}, 085018 (2011).
  
\bibitem{modgravity3}
  F.~S.~N.~Lobo,
  Class.\ Quant.\ Grav.\ {\bf 25}, 175006 (2008).

\bibitem{modgravity4}
  F.~S.~N.~Lobo,
  Phys.\ Rev.\ D  {\bf 75}, 064027 (2007).
  
\bibitem{Lobo:2008sg} 
  F.~S.~N.~Lobo,
  arXiv:0807.1640 [gr-qc].  


\bibitem{Kazanas:1988qa}
  D.~Kazanas and P.~D.~Mannheim,
  Astrophys.\ J.\ Suppl.\  {\bf 76}, 431 (1991).


\end{thebibliography}
\end{document}